# Competing Neural Networks for Robust Control of Nonlinear Systems

Babak Rahmani,[1,*] Damien Loterie,[1] Eirini Kakkava,[2] Navid Borhani,[2] Uğur Teğin,[1,2] Demetri Psaltis,[2] Christophe Moser,[1]

[1]*Ecole Polytechnique Fédérale de Lausanne, Laboratory of Applied Photonics Devices, CH-1015 Lausanne, Switzerland*

[2]*Ecole Polytechnique Fédérale de Lausanne, Laboratory of Optics, CH-1015 Lausanne, Switzerland*

*Corresponding author: babak.rahmani@epfl.ch*

The output of physical systems is often accessible by measurements such as the 3D position of a robotic arm actuated by many actuators or the speckle patterns formed by shining the spot of a laser pointer on a wall. The selection of the input of such a system (actuators and the shape of the laser spot respectively) to obtain a desired output is difficult because it is an ill-posed problem i.e. there are multiple inputs yielding the same output.

Machine learning approaches for imaging have been very successfully implemented in photonics, for such ill-posed problems. For example, the original input phase and amplitude objects can be recovered with 95% fidelity by a neural network from its output intensity diffraction pattern or from the speckle pattern formed after propagation through a scattering medium. The neural network is trained using many different input-output pairs to learn to guess the input from a measurement of the output. This scheme works by having examples of inputs that can produce outputs in the class of output objects the user want produce, but such examples only become available once we have trained the system.

In this paper, we propose an approach that provides a robust solution to this dilemma for any physical system. We show that it is possible to find the appropriate input of a system that results in a desired output, despite the input-output relation being nonlinear and\or with incomplete measurements of the systems variables. We showcase our approach using an extremely ill-posed problem in imaging. We demonstrate the projection of arbitrary shapes through a multimode fiber (MMF) when a sample of intensity-only measurements are taken at the output. We show image projection fidelity as high as ~90 %, which is on par with the gold standard methods which characterize the system fully by phase and amplitude measurements. The generality as well as simplicity of the proposed approach provides a new way of target-oriented control in real-world applications.

# 1. Introduction

Tailoring a physical system to behave in the way that leads to a desired outcome entails careful characterization of the system. The situation becomes worse when the physical system is difficult to model. Even when a model is available, finding the appropriate input that would experimentally result in the desired output is usually not straightforward. Light propagation in scattering media, or through a network of on-chip waveguides are examples of such systems[1-18]. In these situations, the system needs to be characterized by closely monitoring the system's response to a series of inputs and then inversely solving for the system's *response function H*. Often when outputs have a linear relationship with their corresponding inputs and more importantly are fully measured, it is possible to establish a model to relate the inputs to the outputs. Even when a forward response function of a system is found, calculating its inverse $H^{-1}$ could still be challenging because of the large size of H. Fully measuring the outputs of a system requires nontrivial sensory apparatus. For example, the electromagnetic fields satisfying Maxwell equations at the output of a waveguide are complex-valued fields having both amplitude and phase information. Measuring both requires phase extraction using interferometry-based schemes[3]. Being sensitive to environmental perturbation, the phase information needs to be carefully traced and corrected while characterizing the system. The conventional methods cannot provide an optimal solution for a desired output of the system if no phase information is provided.

Neural networks perform well in solving ill-posed inverse problems[19, 20] for various applications such as biology[21], design of photonic devices[22] or novel drugs[23]. Here, we propose a universal neural-network based approach to solve the highly ill-posed problem of predicting a system's *forward and backward* response functions. To this end, we only need a signal modulator to shape the control signal entering the system as well as a detection apparatus to measure the output signal. We show that it is possible to recover the sought-after functions even using the most basic sensory devices that only detect a portion of the outputs' information. The proposed approach is applicable to systems whose inputs and outputs follow a nonlinear relation or even to systems for which the physical mechanism is not known. While potentially applicable to miscellaneous disciplines, in photonics, one can readily envision using the proposed method for light propagation control in spatial and\or spectral domains.

Without loss of generality, we showcase the success of the approach for the scenario of imaging through MMFs: an extreme case for which thousands of modes of the fiber are controlled in order to transmit tens of thousands of pixels though the fiber to project a user defined image. To the best of our knowledge, this is the first demonstration of controlling a nonlinear system with a fidelity on a par with the case for which all parameters of the system are measured (amplitude and phase).

# 2. Results

**Image transmission through Multimode fibers.** The energy splitting among different modes of the multimode fiber (MMF) gives rise to a distorted field at the output and creating speckle patterns[1]. To undo modal scrambling inside the MMF, several approaches ranging from digital iterative algorithms[2-6] to analog\digital phase conjugation[7-13] as well as interferometry methods have been proposed to compensate the modal dispersion in MMFs[14-18]. The former requires calibrating the system for each image to project, which is too computationally intensive and thus too slow for practical applications while the latter entails measuring the complex output field (both phase and amplitude), which is cumbersome to implement. Here, we seek an approach that can be implemented with a simple optical setup, i.e. an input modulator together with an amplitude-only output detection (neglecting phase information), and yet has the generality of approaches that measure both phase and amplitude information. The schematic of such a system is depicted in Fig. 1. Neural networks have previously been shown to reconstruct the undistorted input fields of scattering media including MMFs[24-28] from the amplitude-only scrambled speckle patterns at its output for fibers. Here we intend to accomplish the reverse: to learn the correct inputs that will generate a desired output of the MMF. This is challenging because first and foremost, no straightforward way (without using the transmission matrix) exists for acquiring a training set with the desired output patterns to train the neural network. The training set for this task consists of examples of the desired shapes (for example recognizable drawings such as smiley face, alphabet letters, etc.) on the distal end of the fiber (camera side) and examples of the inputs that generate those patterns when sent through the fiber. Unfortunately, this is exactly the problem we intend to solve. If we had the ability to generate those examples, the problem would already be solved, making it a chicken and egg dilemma. In this regard, an initial attempt towards projecting patterns through MMFs using neural networks has been made that is limited only to focusing spots[29] after the fiber.

**Learning algorithm**. We use a combination of neural networks (referred together as *the projector network*) to generate input control signals that create the desired target output on the detector. Specifically, our projector network, schematically depicted in Fig. 2a, is made of two sub-networks, the *Generator* and the *Discriminator*. The discriminator sub-network (D) tries to learn the forward propagation path of the light through the system (input to output) and the generator (G) learns the inverse path. In other words, G generates proper inputs constrained by the physical propagation rules of the system embedded in the network as D. The two networks are trained synergistically so that D forces G to generate control patterns that, upon sending through the fiber, produce images on the detector belonging to the desired set. Training D to emulate the forward path of the light propagation in fact allows *in situ* back propagation of the error between the desired target images and what appears on the detector in the real experiment through the virtual fiber (i.e. D). When the back propagated error reaches G, the learnable parameters of the sub-network G is adjusted so as to the error is reduced. The error is the smallest when G is the inverse of D. Therefore, this training method, which has also been proposed for use in control systems nearly three decades ago[30], is effectively trying to find the inverse path of light propagation or analogously the time reversal operator.

The architecture of the projector network is *reminiscent* of the so-called conditional Generative Adversarial Networks (GANs). The similarity is rooted from the fact that the training of G cannot be straightforwardly carried out because no label (ground truth control patterns) for target output images exists *a-priori* (training cannot be performed in a supervised manner in which ground truth labels are available beforehand). Therefore, the performance of G gets better by working synergistically with D to generate control patterns that result in output images with higher fidelities. The discriminator network evaluates the patterns produced by the generator based on the fidelity with which the D network mimics the experimental measurements. The training procedure is explained formally in the Materials and Methods section.

**Image projection.** We first train our network with grayscale images of handwritten Latin alphabet from EMNIST[31] as targets (see data preparation subsection in Materials and Methods for more information). Fig.

3 depicts examples of these images and the physical outputs on the camera projected using the proposed learning algorithm. For the sake of comparison, target images are also projected with the transmission matrix approach, an example of methods that require full measurement and control of phase information. In each image, the inset indicates the projection fidelity.

Without any fine tuning, the network trained with Latin characters is used directly to project other category of images. Examples of projected images are shown in Fig. 4, Supplementary Figure 1 and Supplementary Video 1. These results demonstrate the generalization ability of the projector neural network and show that it can extend its ill-posed inverse problem ability to images never seen by the network even in the training step.

Table I summarizes the projection fidelity of the neural network approach and that of the transmission matrix approach for various types of images.

The performance of the neural network in inferring the required SLM modulations is correlated with the complexity of the target images that the network is trained with. The Latin alphabet images, used for training the network in the first experiment, are of sparse nature: having a constant zero background and a grayscale feature centered in the middle of the image. Therefore, it is expected that projecting target images with richer contexts will be more challenging.

To show this, we use our approach to project continuous gray-scale natural-scene-like pictures. Examples of which are shown in Fig. 5. The projected images (red, green, blue, 3-channel RGB and the superposition of all three as one channel) are also depicted. The complexity in the target images makes the training difficult; however, our method is able to provide the appropriate SLM modulations for projecting images of natural scenes with fidelities on par with that of full-measurement schemes.

### 3. Discussion

**Learning trajectory**. The fidelity plot in Fig. 6a demonstrates the algorithm's convergence speed in finding an appropriate solution for the system. The convergence speed depends on the complexity of the target images, the modulation scheme and the extent by which this modulation can be implemented via the modulator, and finally the rate at which the system changes over time. For example, instead of complex-value solutions, the network can be forced to find solutions that are amplitude-only. The former type of solutions is better suited for SLMs and the latter is used more conveniently with amplitude-only modulators (such as digital micromirror devices). It can be shown that the number of iterations required to achieve certain fidelity is higher when an amplitude-only solution is favored (5 iterations versus 1 to 2 in the complex-value case). Refer to the Supplementary Figure 2 for comparing the quality of the projected patterns for amplitude-only inputs. Still, for complex-value solutions, an intermediate conversion scheme is required to convert them into phase-only solutions compatible with phase-only SLMs. On the other hand, the fiber is also prone to multiple time-dependent processes including mechanical perturbations, instabilities associated with drifts in power and working wavelength of the laser source, among others, which influence the learning trajectory.

In a first step to investigate the robustness of the projector algorithm, we use the measured transmission matrix of the system (measured once) to virtually forward the control patterns provided by our algorithm. Doing this, we are able to obtain the resulting fiber's outputs without sending them directly through the fiber. In this way, the necessary intermediate modulation step needed for converting the patterns to a second set of patterns compatible with the modulator is bypassed altogether. The projector network is then trained as before with this new dataset. After training, the network is run to produce the SLM patterns that correspond to a user defined output. The patterns are then loaded on the SLM of the experimental system. The fidelity trajectory of the projected images is shown in Fig. 6c (solid lines) for three colors (Red, Green, Blue). It can be inferred from the plots that the fidelities of the projected images converge to slightly higher values than those of the experimentally projected images (solid circles). The lower fidelity of the latter is because of the degradation due to time variation and non-perfect modulation scheme. The transmission matrix used for forwarding the SLM patterns could also be re-measured after each round of training as to bring the system's variation with time into play (dashed lines in Fig. 6c). Hence although the system is changing over time, it is effectively being corrected. Interestingly, the close overlap between trajectory of graphs in Fig. 6c (dashed lines) and the experimentally projected images (solid circles) shows that the neural network approach is automatically correcting for the drifts but without the need to continuously measure and invert the matrix as it is required in the transmission matrix approach.

**Network's special architecture.** The use of two sub-networks together for generating SLM patterns of the target camera images is better understood when they are replaced entirely with one single network that is trained to predict the fiber's input from its output (guessing SLM images from speckle patterns) (refer to Supplementary Figure 3 for more details). It is shown that the latter network has significantly lower projected image fidelities. Therefore, the superiority of the two sub-network approach is very much indebted to the sub-network D learning the forward path of the optical propagation inside the fiber. This allows the sub-network G to immediately evaluate the projection performance and to adapt accordingly to obtain a better projection *in the course of training* whereas with the single network approach direct evaluation of the projection performance is not possible

### 4. Materials and Methods

*Experimental set-up:* The experimental setup for image transmission through the fiber is depicted in Fig. 7. Three continuous input beams at wavelengths 488, 532, and 633 nm are delivered one at a time to the system via a single mode fiber. The beam entering the system (attenuated to an average power of 4 mW), is collimated by lens L2 (*f*=100 mm) and then directed to the SLM. The beam spatially modulated by the phase-only SLM (HOLOEYE PLUTO) is imaged on the input facet of a multimode fiber using a 4-f system composed of L3 (*f*=250 mm) and OBJ 1 (60x, NA=0.85). After transmission through the graded-index fiber with length L=75 cm, core diameter D=50 μm and a NA of 0.22 (corresponding to ~1050 fiber modes for one polarization), the output field is imaged onto the camera using an identical 4-f configuration.

# Figures

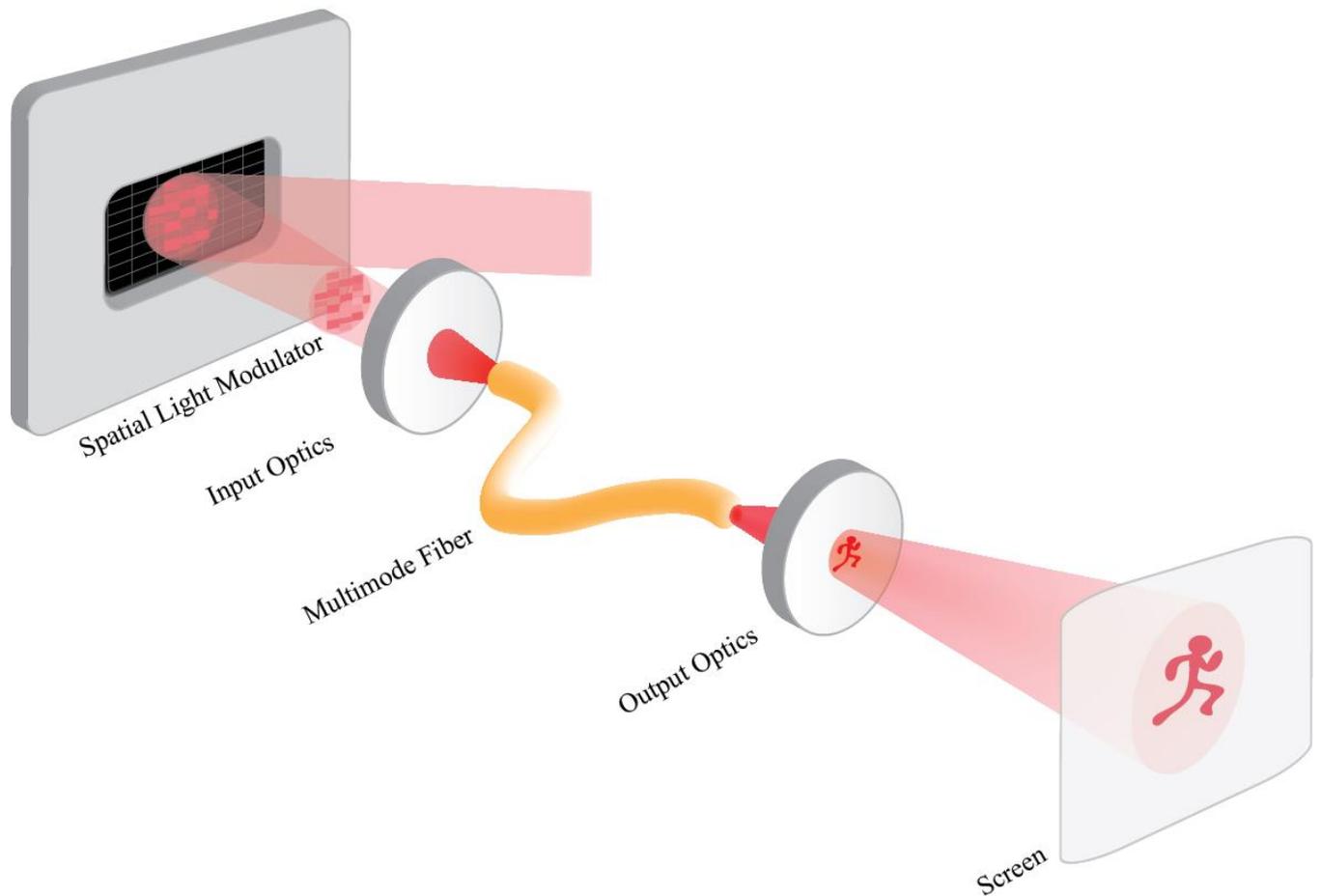

**Figure 1. Fiber projector:** The setup consists of an input signal (laser source), a light modulator (SLM), the system (multimode fiber), and a screen. In the training phase, a detector (camera) is used in place of the screen to record the system's responses to the inputs. The neural network is then fed with the pairs of input signals and the system's responses so that it finds the forward and backward response functions of the system. Once trained, the network is given the target output (the picture of the "running man" here) that is to be projected on the screen and in return it gives out the appropriate control signal that would result in the target image (the network finds the modulation required to undo the light's scrambling inside the fiber).

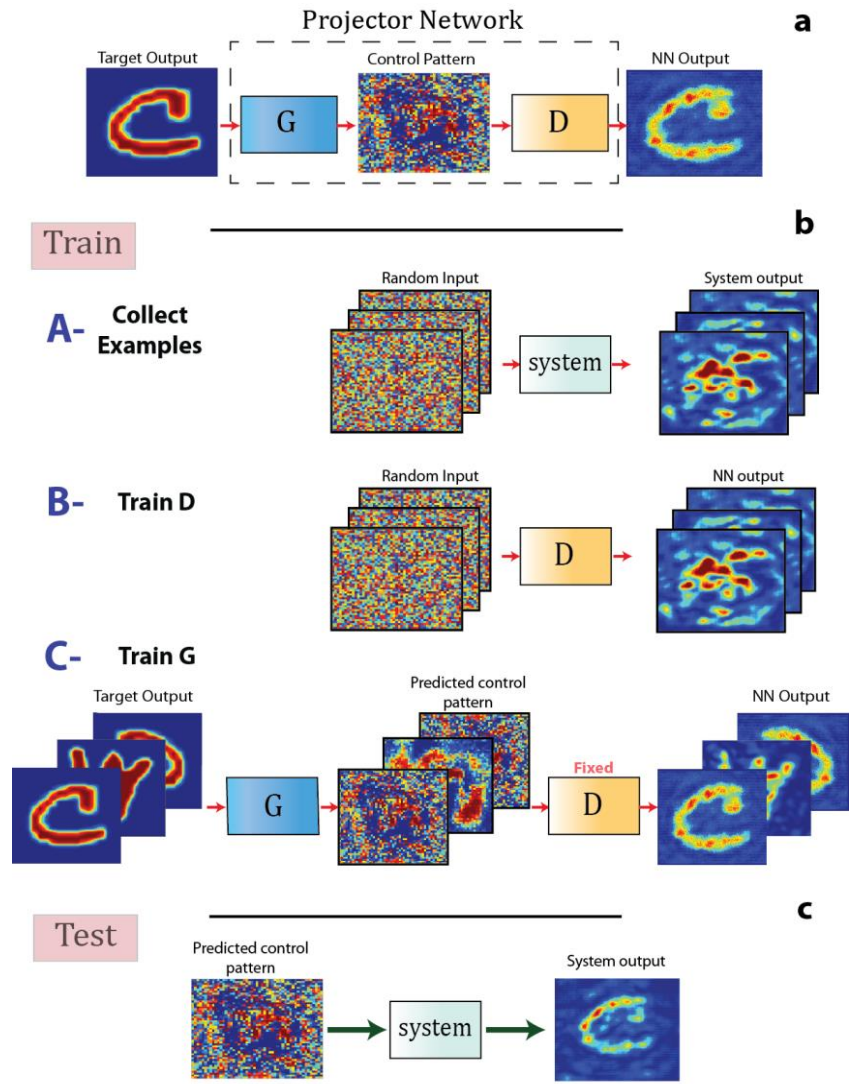

**Figure 2. Neural network's architecture, training and validation procedure: a,** the overall schematic of the projector network consists of two sub-networks: a discriminator (D) and a generator (G). Once trained, the sub-network G accepts a target pattern desired to be projected at the output of the system (here MMF) and accordingly generates a control pattern (here SLM image) corresponding to the target pattern. The role of the sub-network D is to help the network G to come up with control patterns that are bound by the physics of light propagation through the fiber. **b,** the training procedure is carried out in three steps: **A-** a number of input control patterns are sent through the system and the corresponding outputs are captured on the camera. **B-** the sub-network D is trained on these images to learn the mapping from the SLM to camera; hence D is essentially learning the optical forward path of light starting from its reflection from the SLM, propagation through the MMF and finally impinging on the camera. **C-** while sub-network D being fixed, G is fed with a target image and is asked to produce an SLM image corresponding to that target image. The G-produced SLM image is then passed to the fixed sub-network D now mimicking the fiber. The error between the output of D and the target image is back propagated via D to G to update its trainable weights and biases. **c,** the test procedure is carried out by feeding the target image to the trained sub-network G and acquiring the appropriate SLM image corresponding to that target image and sending it through the system.

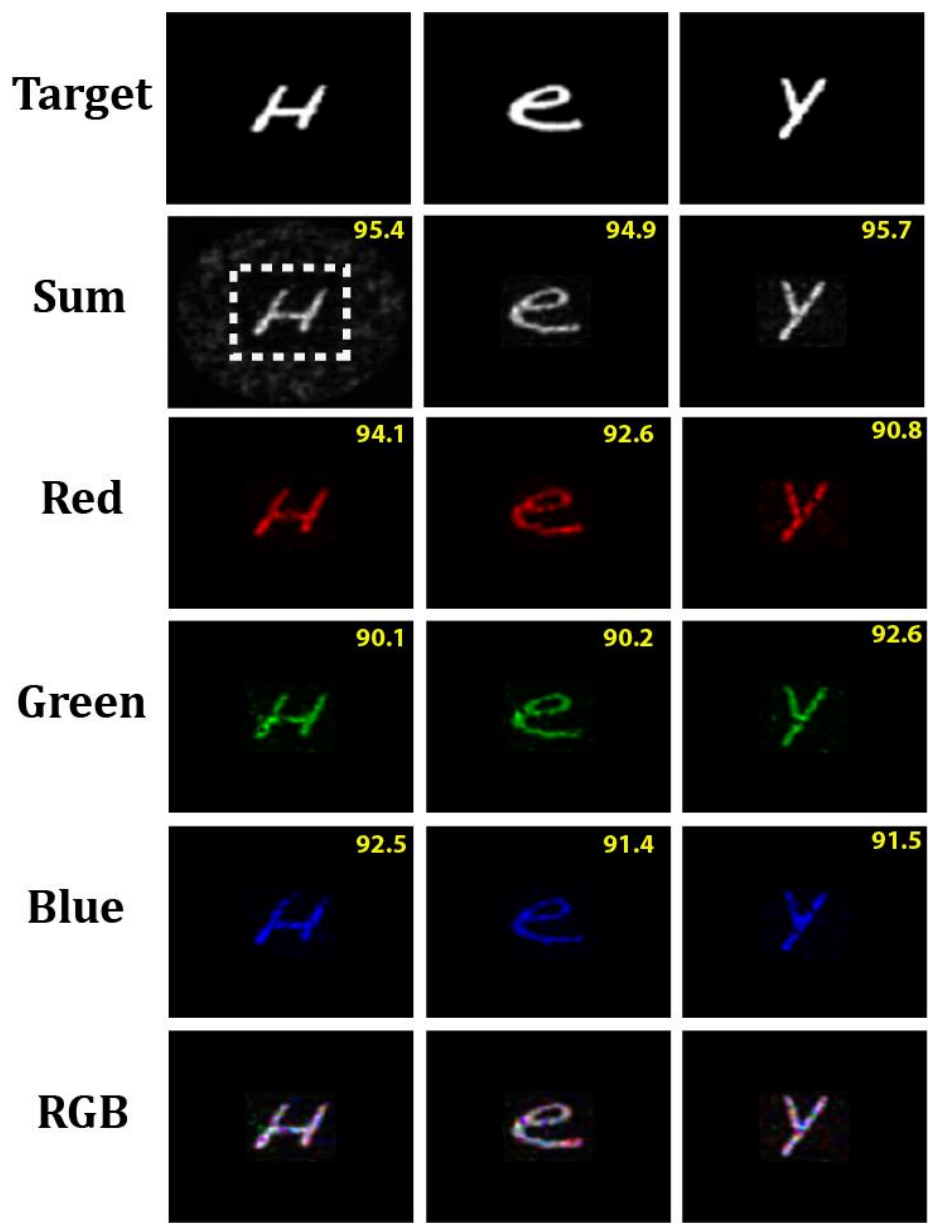

**Figure 3. MMF system arbitrary output control with partial measurements:** Examples of images projected onto a camera at the output of a MMF are shown. The projection of images are carried out for three different wavelengths (633 nm, 532 nm, 488 nm) corresponding to red (R), green (G) and blue (B) as well as the superposition of those colors either as a 3 channel RGB image or as a one channel incoherent image produced by summing R, G and B. The neural network is trained with EMNIST dataset as target images. The appropriate SLM patterns generated by the network are sent to the system to obtain the desired targets on a rectangular area of size 200×200 pixels on the camera (corresponding to an area of 19×19 μm$^2$ on the output facet of the fiber). This area is shown as a dashed box on one of the examples. The fidelity of projected images with respect to the corresponding target images is also shown.

**Figure 4. Neural network generalization ability for controlling the MMF output:** Examples of images projected onto a camera at the output of a MMF are shown. The control patterns that produce the output images on the camera (the incoherent summation of red, green and blue wavelengths as well as the 3 channel RGB images) are generated either via a neural network (NN) trained on the dataset of Latin alphabet characters (different from the category of target images) or via the transmission matrix full measurement approach (TM). The generalization of the network is demonstrated in its ability to provide control patterns for target images that come from a different class as that of the images originally used for training.

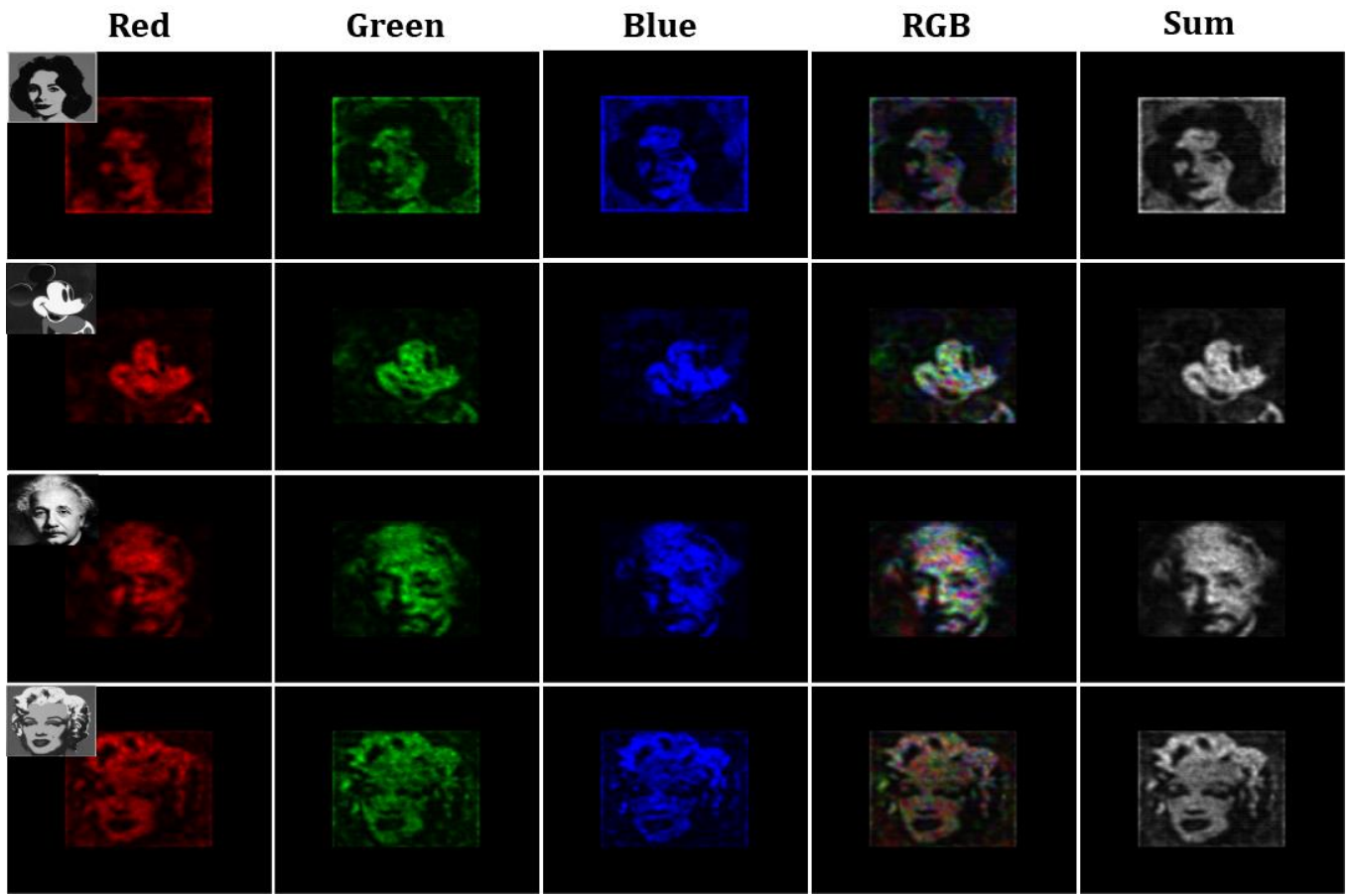

**Figure 5. Continuous gray-scale image projection**: Examples of natural-scene continuous gray-scale target and experimentally projected images being sent through the MMF and captured on the camera for colors red, green, blue and the 3-channel RGB as well as the superposition of all three colors in 1 channel (sum) are shown.

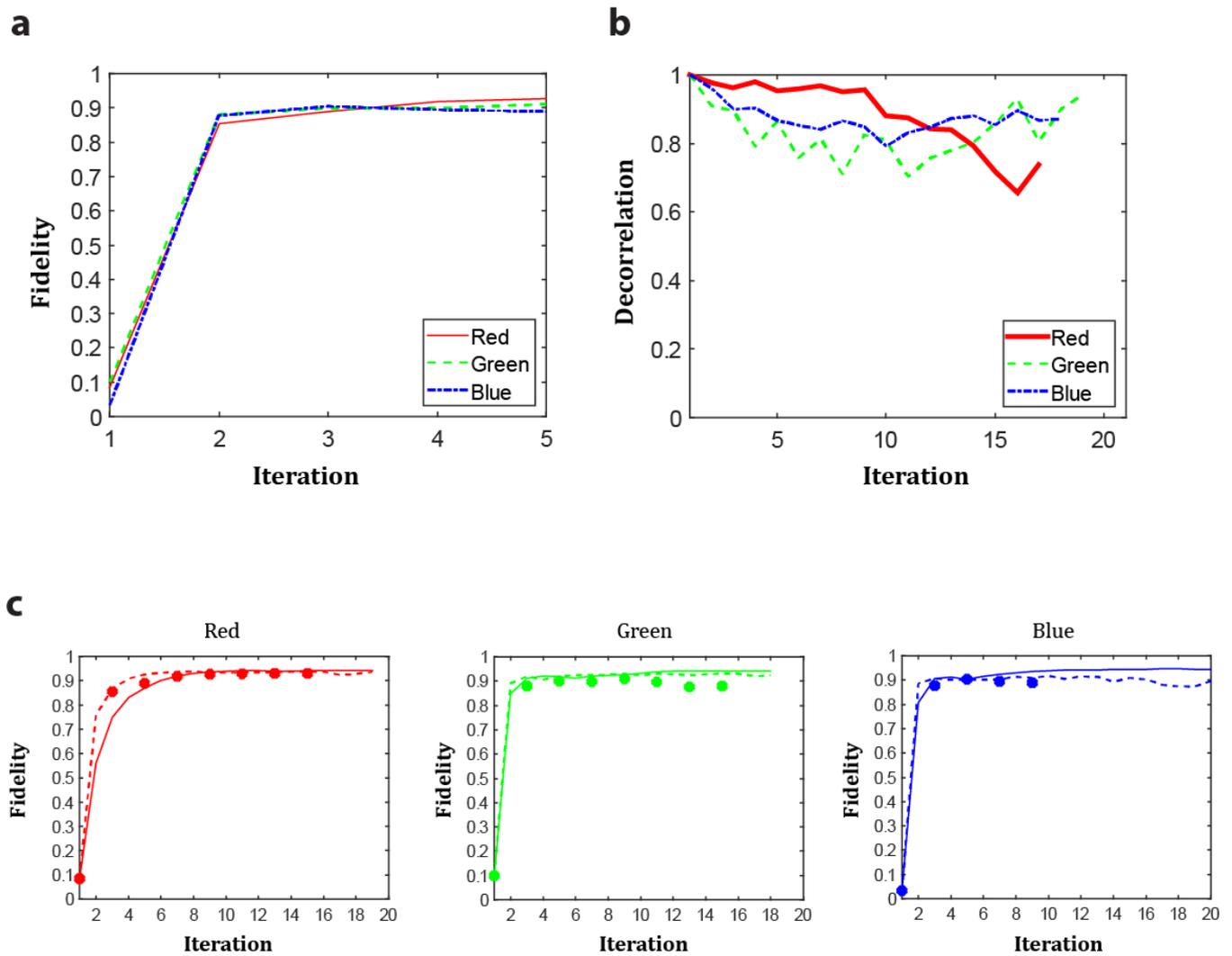

**Figure 6. Fidelity trajectory of the system's output**. **a**, the fidelity trajectory of experimentally projected images versus the training iteration number is plotted for all three colors. **b**, while training, the instability of the system (estimated as the correlation between instances of the fiber's response to a constant input signal) is monitored over time. **c**, degradation in the fidelity of projected images due to the non-perfect modulation scheme as well as the variation of the system with time is shown by using the experimentally measured transmission matrix (TM) to forward the neural network's predicted SLM images for all three colors. The fidelities in part (**a**) are redrawn in part (**c**) for comparison. As observed, the experimentally projected images (solid circles) closely follow the track of time variant TM-based relayed projections (dashed lines) and both eventually fall below the track of time-invariant TM-based relayed projections (solid lines). In the former, what is taken out from the learning algorithm is only the effect of modulation scheme, whereas in the latter, it is the lumped effect of time variation as well as the modulation scheme. The ripples in the trajectory of the graphs in (**c**) (dashed lines) show that the network is continuously trying to correct for the drifts.

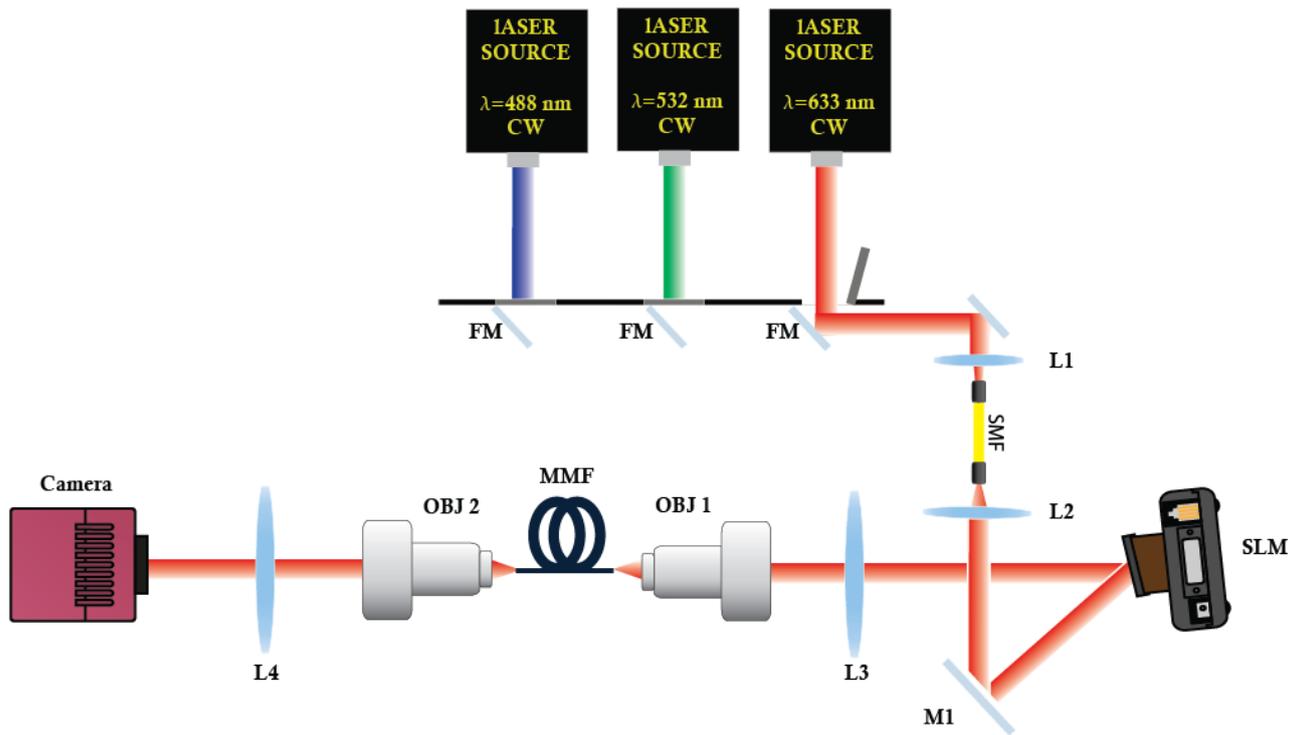

**Figure 7. Optical setup**: Detailed diagram of the optical setup. Control patterns are generated via the SLM, guided through the fiber and captured by the camera. L1: Aspheric lens, L2: f = 100mm lens; L3: f = 250mm lens; L4: f = 250mm lens; OBJ1, OBJ2: 60x microscope objective; SLM: spatial light modulator; M1: mirror; FM: flip mirror; SMF: single mode fiber; MMF: multimode fiber.

**Table 1** Neural network and transmission matrix image projection average fidelities (in percent) for various dataset

| Dataset (1000 samples) | Superposition of all three wavelengths | | Red | | Green | | Blue | |
|---|---|---|---|---|---|---|---|---|
| | NN | TM | NN | TM | NN | TM | NN | TM |
| Latin alphabet | 95.03 | 98.09 | 92.54 | 96.01 | 90.73 | 96.64 | 91.22 | 97.22 |
| Digits | 95.16 | 98.15 | 92.38 | 96.24 | 90.92 | 96.68 | 91.59 | 97.30 |
| Random sketches | 86.62 | 91.17 | 83.26 | 88.96 | 83.57 | 89.76 | 81.96 | 91.15 |